\input harvmac
\def\half{{1 \over 2}}

\def\>{{\rangle}}
\def\<{{\langle}}

\def\p{{\partial}}

\def\s{{\sigma}}

\def\L{{\Lambda}}
\def\l{{\lambda}}

\def\a {{\alpha}}
\def\b {{\beta}}
\def\g {{\gamma}}

\def\G {{\Gamma}}
\def\d {{\delta}}

\def\e {{\epsilon}}

\Title{\vbox{\hbox{IFT-P.038/97}}}
{\vbox{\centerline{\bf Ramond-Ramond Central Charges in the}
\centerline{\bf Supersymmetry Algebra of the Superstring}}}
\bigskip\centerline{Nathan Berkovits}
\bigskip\centerline{Instituto
de F\'{\i}sica Te\'orica, Univ. Estadual Paulista}
\centerline{Rua Pamplona 145, S\~ao Paulo, SP 01405-900, BRASIL}
\bigskip\centerline{e-mail: nberkovi@ift.unesp.br}
\vskip .2in
The free action for the massless sector of the Type II superstring was
recently constructed using closed RNS superstring field theory.
The supersymmetry transformations of this action are shown
to satisfy an N=2 D=10 SUSY algebra with Ramond-Ramond central charges. 

\Date{June 1997}
\newsec {Introduction}

One of the most important reasons for believing superstring theory is
related to a higher-dimensional theory is the fact that non-perturbative
superstring states called $D$-branes carry Ramond-Ramond charge\ref\Pol
{J. Polchinski, Phys. Rev. Lett. 75 (1995) 4724.}. This provides a connection
between eleven-dimensional
Kaluza-Klein states and superstring theory since Ramond-Ramond 
charge is related to momentum in the eleventh direction.
In this letter, it is shown that R-R central charges are present in the
supersymmetry algebra of the superstring, providing an even more direct
connection with a higher-dimensional theory.

Ramond-Ramond central charges are usually assumed to be absent in the
spacetime-supersymmetry algebra of the superstring. This assumption is
based on three arguments: 1) No perturbative superstring states carry
Ramond-Ramond charge; 2) The left and right-moving spacetime-supersymmetry
generators anti-commute with each other in the RNS formalism;\foot{The
left-moving with left-moving SUSY generators anti-commute to give the
usual momentum and NS-NS one-brane charge, but no NS-NS five-brane 
charge.\ref
\Poly{D. Polyakov, `A Comment on a Fivebrane Term in Superalgebra'
(Revised version), hep-th 9704191.}} 
and 3) There is
no Wess-Zumino 
term for Ramond-Ramond fields in the standard GS sigma model.\ref\az
{J.A. de Azcarraga, J.P. Gauntlett, J.M. Izquierdo and P.K. Townsend, 
Phys. Rev. Lett. 63 (1989) 2443.}

Although all of these arguments are correct, it will be shown here that
R-R central charges are nevertheless present in the SUSY algebra of the
superstring. Because of the first argument, these charges can be detected
only through the abelian transformation of the R-R gauge field.
Note that when ``picture'' is treated
properly \ref\sft{N. Berkovits, Phys. Lett. B388 (1996) 743, 
hep-th 9607070\semi
N. Berkovits, Phys. Lett. B395 (1997) 28, 
hep-th 9610134.},
zero-momentum R-R gauge fields are present in the string field,
although they decouple from all perturbative superstring states. Using
superstring field theory, the transformation of the R-R gauge field is enough
to construct the global R-R charge out of RNS matter and ghost worldsheet
variables. When acting on
perturbative superstring states, this charge is BRST-trivial, which is
not ruled out by the second argument since the supersymmetry algebra in the
RNS description 
closes only up to gauge transformations and equations of motion.
Finally, the dependence of the R-R charge on worldsheet ghosts
explains why the third argument is inapplicable. In the standard GS sigma
model, worldsheet ghosts are not understood, as can be seen by the absence
of a Fradkin-Tseytlin term\ref\frad{E.S. Fradkin and A.A Tseytlin,
Phys. Lett. B158 (1985) 316.} which should couple the spacetime dilaton to
worldsheet curvature.

Because of picture-changing difficulties, a free action for the R-R fields of
the Type II superstring has only recently been constructed\sft.
In the massless
sector, the R-R string field depends on an ``electric'' gauge field
$A_{(+)}^{\a\b}$, a ``magnetic'' gauge field 
$A^{\a\b}_{(-)}$, and an infinite
set of auxiliary fields $F_{(n)}^{\a\b}$ for $n$=1 to $\infty$.
($\a$ and $\b$ are 16-component SO(9,1) spinor indices whose chirality
depends if one is discussing Type IIA or Type IIB.) 
This set of massless fields can be understood using the
usual RNS formalism if the R-R vacuum state is chosen to be annihilated
by $\g_L^0$ and $\b_R^0$ (as opposed to $\b_L^0$ and $\b_R^0$) where
$[\b_L^0, \g_L^0]$ and
$[\b_R^0, \g_R^0]$ are the zero modes of the
left and right-moving bosonic ghosts.

After adding the NS-R, R-NS, and NS-NS massless sectors, 
the action is invariant
under coordinate and $b_{\mu\nu}$ gauge transformations, under local
N=2 D=10 supersymmetry transformations, and under R-R gauge transformations
of $A_{(+)}^{\a\b}$ and 
$A_{(-)}^{\a\b}$. It will be shown here that the
commutator of two supersymmetry transformations gives a coordinate
transformation, a $b_{\mu\nu}$ gauge 
transformation, and a R-R gauge transformation. This means that the 
SUSY algebra of the Type II superstring contains R-R central charges
as well as the usual NS-NS one-brane central charge. 

Note that even without knowing superstring field theory, the 
presence of Ramond-Ramond central charges could have been predicted
based on the role of the dilaton as a conformal compensator\ref\comp
{N. Berkovits and W. Siegel, Nucl. Phys. B462 (1996) 213, hep-th 9510106.}.
In N=2 D=4 supergravity coupled to vector multiplets, it is well-known
that the N=2 D=4
SUSY algebra contains central charges of the form\ref\dewit
{B. deWit, P.G. Lauwers and A. Van Proeyen, Nucl. Phys. B255 (1985) 569.} 
\eqn\dew{[ \d_{q^\a_i} (u^\a_i) ,
\d_{q^\b_j} (v^\b_j) ] = 
\sum_{n} \d_{C_{(n)}} ( W_{(n)} u^{i\a} v^j_\a \e_{ij} )}
where $i$ is an SU(2) index, $W_{(n)}$ is the expectation value of the
complex scalar in the $n^{th}$ vector multiplet, 
$\d_{q^\a_i} (u^\a_i)$ is the SUSY transformation with parameter 
$u^\a_i$, and $\d_{C_{(n)}} (\rho_{(n)})$ is the gauge transformation of
the $n^{th}$ gauge field with parameter $\rho_{(n)}$.

One of these vector multiplets is the vector compensator multiplet and,
as shown in \comp, the expectation value of the scalar in this multiplet
is $<e^{-\phi}> = g^{-1}$ where $\phi$ is the 
dilaton and $g$ is the string coupling constant.
So one expects R-R central charges in the Type II superstring, at least
after compactifying to four dimensions.\foot{The $g^{-1}$ factor
in this central charge can be explained by recalling that the
gauge field in the conformal compensator, $\hat A^{\mu}$, is related to
the R-R gauge field in superstring field thoery, $A^\mu$, by
$\hat A^\mu =e^{-\phi} A^\mu$. (This can be seen from the fact that
the kinetic term for $\hat A_\mu$
carries no $e^{-2\phi}$ factor in the
tree-level effective action.) 
For this reason, there is no $g^{-1}$ factor in the
SUSY algebra when R-R charge is defined with respect
to $A^\mu$ rather than $\hat A^\mu$.} 
This suggests that a proper
understanding of conformal compensators in ten dimensions will help
in understanding the role of R-R central charges. 

\newsec {Superstring Field Theory}

\subsec{The massless R-R sector}

The massless R-R contribution to the free Type II superstring
action was
recently constructed using closed superstring field theory.\sft
Although ``non-minimal'' fields\ref\Sieg
{W. Siegel, Int. J. Mod. Phys. A6 (1991) 3997\semi
N. Berkovits, M.T. Hatsuda and W. Siegel, Nucl. Phys. B371 (1991) 434.} were
needed for this construction, the equations of motion and gauge invariances
can be easily analyzed using the standard RNS worldsheet variables.

Physical closed superstring states are described by fields $|\Phi \>$ 
of zero total
ghost-number 
where total ghost number is $g_L +g_R -2$, $g_{L/R} = \oint d\s
(b_{L/R} c_{L/R} -
\eta_{L/R}\xi_{L/R})$,
and the bosonic ghosts have been fermionized as $\beta=\p\xi e^{-\phi}$ and
$\gamma=\eta\e^{\phi}$. (The $-2$ is present so that 
physical states carry zero ghost number, implying that
the SL(2)-invariant vacuum
carries $-2$ total ghost-number.)
Note that this definition of ghost-number is slightly modified from the
standard one, $g =\oint d\s(b c -\p\phi)$, 
although they agree at zero picture.
(Picture is defined by $P=\oint d\s (\eta\xi -\p\phi)$.)
The modified definition of $g$ is necessary in order that the
spacetime-supersymmetry generators carry zero ghost number.

These string fields must satisfy the constraints $b^0_- |\Phi \>=
L^0_- |\Phi \>=\eta^0_{L/R} |\Phi \>= 0$ and
are defined up to the gauge transformation $\d |\Phi \>= Q |\L \>$
where $Q=Q_L+Q_R$ is the BRST charge, $b^0_-= b^0_L - b^0_R$, $L^0_-=
L_L^0 -L_R^0$ is the difference of the left and right-moving energies,
$\eta^0_{L/R} |\Phi\>=0$ implies no dependence on $\xi_{L/R}^0$,
and $|\L\>$ is a gauge field of $-1$ total ghost-number satisfying
$L^0_- |\L\>  =b^0_- |\L\>=\eta^0_{L/R} |\L\>=0$. 
The equation of motion is $Q|\Phi\>=0$.

Finally, there is a constraint coming from the restriction to a single
picture. Surprisingly, different choices of left and right-moving picture,
$P_L$ and $P_R$, give different sets of on-shell fields.\foot{The usual
argument for picture independence of BRST cohomology comes from the
fact that $Y Z=1$ where $Y= c\p\xi e^{-2\phi}$, $Z =\{ Q, \xi\}$,
and $\p Y$, $\p Z$ are BRST trivial. 
For the open superstring, $Q |V\>=0$ and $|V\> \neq Q |\L\>$ implies that
$Q Z |V\> =0$ and $Z |V\> \neq Q |\hat\L\>$ (since 
$Z |V\> =Q |\hat\L\>$ implies $|V\> = Q Y |\hat \L\>$).
But for the closed superstring, 
$Q |V\>=0$ and $|V\> \neq Q |\L\>$ does not imply
$Q Z_L |V\> =0$ and $Z_L |V\> \neq Q |\hat\L\>$. This is because
$Y_L$ and $b_-$ do not commute, so although
$Z_L |V\> =Q |\hat\L\>$ implies $|V\> = Q Y_L |\hat \L\>$,  
$b_-^0 Y_L |\hat\L\>$ is not necessarily
zero even when $b_-^0 |\hat\L\>=0$.
Therefore, closed superstring BRST cohomology can
depend on the choice of picture.}
For example,
it will be found that there are on-shell constant modes in the R-R sector if
$(P_L, P_R)= (- 3/2,-1/2)$, but not if $(P_L, P_R)= (-1/2, -1/2)$.
Since one wants to reproduce the equations of motion from reference \sft
where on-shell zero-momentum
R-R fields were found using an action principle,
$(P_L, P_R)$ will be chosen to be $(-3/2, -1/2)$ in the R-R sector.

This means that R-R states can be constructed using non-negative modes
acting on a vacuum $|0\>_{R-R}^{\a\b}$ which satisfies
\eqn\one{b_L^0 |0\>_{R-R}^{\a\b}
=b_R^0 |0\>_{R-R}^{\a\b}=
\g_L^0 |0\>_{R-R}^{\a\b}
=\b_R^0 |0\>_{R-R}^{\a\b}=0.} 
(The picture $(P_L,P_R)=(-\half, -\half)$ would correspond to a vacuum 
satisfying
$\b_L^0 |0\>_{R-R}^{\a\b}=
\b_R^0 |0\>_{R-R}^{\a\b}=0$.) 
In terms of the SL(2)-invariant vacuum, 
$|0\>_{R-R}^{\a\b}$ is $c_L e^{-{3 \over 2}\phi_L}\Sigma_L^\a$
$ c_R e^{-\half\phi_R} \Sigma_R^\b$
where $\Sigma_{L/R}^{\a}$ is the left and right-moving spin field of
weight $5/8$.
The $\psi_L^\mu$ and $\psi_R^\mu$ zero modes are treated like
SO(9,1) gamma-matrices which transform the bispinor indices on 
$|0\>^{\a\b}_{R-R}$. Furthermore, all string states 
must be GSO-projected in the
usual way.

Constructing all possible states of zero ghost-number, the massless
R-R states of the Type II superstring are given by
\eqn\two{
|\Phi\>_{R-R}
= \sum_{n=0}^\infty (A_{(n)}^{\a\b}(x) (\b_L^0)^n (\g_R^0)^{n}
+
F_{(n)}^{\a\b}(x) c_+^0 (\b_L^0)^{n+1}(\g_R^0)^{n})~ |0\>_{R-R}^{\a\b}.}
where $c_+^0 = c_L^0 + c_R^0$.
Note that the GSO projection implies that the bispinor indices on
$A_{(n)}^{\a\b}$
and $F_{(n)}^{\a\b}$ have different SO(9,1) chirality if $n$ is even or odd.
(In the above expressions, contracted spinor
indices always have opposite SO(9,1) chiralities).

Using the gauge string field,
\eqn\th{|\L\>_{R-R} = \sum_{n=2}^\infty \L_{(n)}^{\a\b}(x) 
c_+^0 (\b_L^0)^n (\g_R^0)^{n-2}|0\>_{R-R}^{\a\b},}
it is easy to check that $A_{(n)}^{\a\b}$ can be 
algebraically gauged away for $n>1$. The equation 
of motion $Q |\Phi\>=0$, together with the requirement
that only a finite number of fields are non-zero\foot{This requirement
can be interpreted as a normalization condition on the string field\sft.},
implies that the remaining fields 
satisfy\sft
\eqn\mot{\hat F_{(0)}^{\a\b}=\p^{\a\g} A_{(0)}^{\g\b} -
\p^{\b\g} A_{(1)}^{\a\g}, \quad
\p^{\b\g} A_{(0)}^{\a\g} -
\p^{\a\g} A_{(1)}^{\g\b}=0, }
$$ \p^{\a\g}\hat F_{(0)}^{\g\b}=
\p^{\b\g}\hat F_{(0)}^{\a\g}=0,\quad
F_{(2n+1)}^{\a\b}=
F_{(2n+2)}^{\a\b}=0$$
where
$\hat F_{(0)}^{\a\b}=
F_{(0)}^{\a\b} +\p^{\a\g}
A_{(0)}^{\g\b}$ and
$\p^{\a\b}= \G_\mu^{\a\b}\p^\mu$. 
The fields and equations of \mot are the same as those of reference \sft and
correspond to the Bianchi identities and equations of motion
for a zero-form, two-form and four-form field strength (for Type IIA)
or a one-form, three-form and self-dual five-form field strength 
(for Type IIB).

The R-R fields $A_{(\pm)}^{\a\b} 
=A_{(0)}^{\a\b} \pm
A_{(1)}^{\a\b}$ play the role of ``electric'' and ``magnetic'' gauge fields
with the gauge transformations,
\eqn\gau{\d A_{(+)}^{\a\b} =\p^{\a\g}\rho_{(+)}^{\g\b} +
\p^{\b\g}\rho_{(+)}^{\a\g}\quad
\d A_{(-)}^{\a\b}=\p^{\a\g}\rho_{(-)}^{\g\b} -
\p^{\b\g}\rho_{(-)}^{\a\g},} 
described by the gauge string field
\eqn\gsf{|\L\>_{R-R} = (\rho_{(0)}^{\a\b}(x) 
\b_L^0
+\rho_{(1)}^{\a\b}(x) 
(\b_L^0)^2 \g_R^0 -\p^{\b\g}\rho_{(1)}^{\a\g}(x) c_+^0 (\b_L^0)^2)~
|0\>_{R-R}^{\a\b}}
where $\rho_{(\pm)}^{\a\b} 
=\rho_{(0)}^{\a\b} \pm
\rho_{(1)}^{\a\b}$.
The constant modes of $A^{\a\b}_{(\pm)}$
are in the BRST cohomology since $|0\>^{\a\b}_{R-R}$
and $\b_L^0 \g_R^0$ 
$|0\>^{\a\b}_{R-R}$ can not be written as $Q|\L\>$
if $|\L\>$ is restricted to contain a finite number of
terms. 
(If an infinite number of terms
are allowed, $|0\>^{\a\b}_{R-R}=$
$Q  \sum_{n=0}^\infty $
$(-1)^n c_+^0  
(\b_L^0)^{2+2n}(\g_R^0)^{2n}$
$|0\>^{\a\b}_{R-R}$ and 
$\b_L^0 \g_R^0 |0\>^{\a\b}_{R-R}=$
$Q  \sum_{n=0}^\infty$
$ (-1)^n c_+^0$
$ 
(\b_L^0)^{3+2n}$
$(\g_R^0)^{1+2n} |0\>^{\a\b}_{R-R}
$.) One way to see the necessity of this 
restriction on $|\L\>$ is to note that
$\<D|0\>^{\a\b}_{R-R}$ is non-zero where $\<D|$ is the $D$-brane
boundary state\ref\sagn{M. Bianchi, G. Pradisi and
A. Sagnotti, Nucl. Phys. B376 (1992) 365.}. Since $\<D| Q=0$, 
$|0\>^{\a\b}_{R-R}$ can not be BRST-trivial.\ref\san{S. Ramgoolam, private
communication.}

Note that these constant modes of $A^{\a\b}_{(\pm)}$
would not be present if one had
instead chosen the R-R vacuum to be annihilated by $\b_L^0$ and $\b_R^0$.
In such a picture, the only massless R-R state would be $F^{\a\b}$
with the equations of motion $\p^{\a\g} F^{\g\b}=\p^{\b\g} F^{\a\g}=0$.
However, in this case, 
there is no action in terms of just $F^{\a\b}$
(i.e. without gauge fields)
which can reproduce these equations of motion.
This is especially problematic for the Type IIB superstring where
one needs an infinite number of auxiliary fields to write an action 
for the self-dual five-form field strength.\sft\ref\beng
{B. McClain, Y.S. Wu, and
F. Yu, Nucl. Phys. B343 (1990) 689\semi
C. Wotzasek, Phys. Rev. Lett. 66 (1991) 129\semi
I. Bengtsson and A. Kleppe, ``On chiral P-forms'', hep-th 9609102.}

\subsec{The massless NS-NS, NS-R and R-NS sectors}

The NS-NS, NS-R, and R-NS sectors of the Type II superstring are similarly
described by a string field constructed from non-negative modes acting
on a vacuum. The vacuum $|0\>_{NS-NS}$ will be defined in the picture
$(P_L,P_R)=(-1,-1)$, 
$|0\>_{NS-R}^\a$ will be defined in the picture
$(P_L,P_R)=(-1,-\half)$, 
and $|0\>_{R-NS}^\a$ will be defined in the picture
$(P_L,P_R)=(-{3\over 2},-1)$. So like $|0\>_{R-R}^{\a\b}$, these vacua are 
annihilated by all negative modes and by $\b_R^0$, $\g_L^0$ and $b_{L/R}^0$.
In terms of the SL(2) invariant vacuum,
these vacua are $c_L e^{-\phi_L} c_R e^{-\phi_R}$,
$c_L e^{-\phi_L} c_R e^{-\half\phi_R}$,
and $c_L e^{-{3\over 2}\phi_L} c_R e^{-{\phi_R}}$.

The massless states in these sectors are described by
\eqn\sect{|\Phi\>_{NS-NS}=
((g_{\mu\nu}(x) +b_{\mu\nu}(x))\psi_L^{\half \mu}
\psi_R^{\half \nu} +\phi(x)\b_L^\half \g_R^\half +\tilde\phi(x)
\g_L^\half \b_R^\half }
$$+ B_\mu(x) c_+^0 \b_L^\half \psi_R^{\half\mu}
+\tilde B_\mu (x) c_+^0 \psi_L^{\half \mu}\b_R^\half)~ |0\>_{NS-NS},$$
$$|\Phi\>_{NS-R}
=(\chi_\mu^\a(x) \psi_L^{\half\mu} +\tau^\a(x) c_+^0 \b_L^\half
+\Xi^\a(x)\b_L^\half \g_R^0)~ |0\>_{NS-R}^\a,$$
$$|\Phi\>_{R-NS}
=(\tilde\chi_\mu^\a(x) c_+^0\b_L^0\psi_R^{\half\mu} +
\tilde\tau^\a(x) c_+^0 \b_R^\half
+\tilde\Xi^\a(x)c_+^0(\b_L^0)^2 \g_R^\half $$
$$+
M_\mu^\a(x) \psi_R^{\half\mu} +N^\a(x)\b_L^0 \g_R^\half)~ |0\>_{R-NS}^\a.$$
Note that [$B^\mu$, $\tilde B^\mu$, $\tau^\a$, $\tilde\tau^\a$] are
auxiliary fields, and [$\tilde\phi$, $M_\mu^\a$ $N^\a$] can be elimiminated
by algebraic gauge transformations.

The remaining fields transform under coordinate reparameterizations,
$b_{\mu\nu}$ gauge transformations, and N=2 local supersymmetry 
transformations as
\eqn\rem{\d (g_{\mu\nu} +b_{\mu\nu})= \p_\mu y_\nu +\p_\nu \tilde y_\mu,\quad
\d\phi =\p^\mu(y_\mu+\tilde y_\mu),}
$$\d \chi_\mu^\a =\p_\mu \e^\a,\quad
\d \tilde\chi_\mu^\a =\p_\mu \tilde\e^\a,\quad
\d \Xi^\a =\p^{\a\b} \e^\b,\quad
\d \tilde\Xi^\a =\p^{\a\b} \tilde\e^\b,$$
which are parameterized by the gauge field
\eqn\ar{|\L\>_{NS-NS}= (y_\mu(x) \b_L^\half \psi_R^{\half\mu}+
\tilde y_\mu(x) \psi_L^{\half\mu}\b_R^\half -
\p^\mu \tilde y_\mu c_+^0 \b_L^\half \b_R^\half)~ |0\>_{NS-NS},}
$$|\L\>_{NS-R}=
 \e^\a(x) \b_L^\half ~|0\>_{NS-R}^\a ,$$
$$
|\L\>_{R-NS}=
 \tilde\e^\a(x)(c_+^0\b_L^0 \b_R^\half
- c_+^0 (\b_L^0)^3 \g_R^\half)~|0\>_{R-NS}^\a .$$
Note that these transformations have been chosen to preserve the
gauge $\tilde\phi=M_\mu^\a=N^\a=0$.

\newsec{Conserved Charges}

When the gauge parameters $[y_\mu,\tilde y_\mu,\e^\a,\tilde \e^\a,
\rho_{(0)}^{\a\b},
\rho_{(1)}^{\a\b}]$ of equations \ar and \th are constants, 
$Q|\L\>=0$. The $-1$ ghost-number states described by $|\L\>$ are
\eqn\char{P_R^\mu=~\b_L^\half \psi_R^{\half\mu} ~|0\>_{NS-NS},\quad
P_L^\mu=~ \psi_L^{\half\mu}\b_R^\half ~ |0\>_{NS-NS}}
$$q_R^\a=~\b_L^\half |0\>_{NS-R}^\a,  
\quad q_L^\a=~(c_+^0\b_L^0 \b_R^\half
- c_+^0 (\b_L^0)^3 \g_R^\half)~|0\>_{R-NS}^\a,$$
$$ C_{(0)}^{\a\b}=~\b_L^0 ~
|0\>_{R-R}^{\a\b}, \quad 
C_{(1)}^{\a\b}=~
(\b_L^0)^2 \g_R^0 ~|0\>_{R-R}^{\a\b},$$
which are
the conserved charges associated with the global part of the gauge
transformations. These states cannot
be written as $Q|\Omega\>$ where $|\Omega\>$ has a finite number
of terms, so they are in the BRST cohomology at
$-1$ ghost number. 

Note that the conserved R-R charges, $C_{(0)}^{\a\b}$ and
$C_{(1)}^{\a\b}$, are in the same cohomology class
as the states $(\b_L^0)^n (\g_R^0)^{n-1} |0\>_{R-R}^{\a\b}$ where $n$
is an arbitrarily large odd or even number. 
This means that the R-R charges act as BRST-trivial operators 
on any superstring state which carries finite left-moving ghost number. 
This includes all perturbative superstring states, but not
$D$-brane boundary states which contain the ghost dependence
$e^{\b_L^0 \g_R^0}$.\ref\yost{S.A. Yost, Nucl. Phys. B321 (1989) 629.}
For this reason, $D$-branes can carry non-zero
R-R charge but perturbative superstring states cannot.
Since R-R charge is related to momentum in the eleventh direction,
there is a relation between non-trivial dependence on worldsheet ghosts
and non-trivial dependence on the eleventh direction.

By hitting with the left and right-moving picture-changing
operators, $Z_L Z_R$, it is easy to see that
the $NS-NS$ charges are the $(P_L,P_R)=(-1,-1)$ versions of
the translation 
and NS-NS one-brane
generators
$\int d\s (\p_\tau x^\mu \pm \p_\sigma x^\mu)$, and
the $NS-R$ and $R-NS$ charges are the
$(-3/2,-1)$ and $(-1,-1/2)$ versions of the
N=2 supersymmetry generators
$q_{L/R}^\a=\int d\s e^{-\half\phi_{L/R}}\Sigma^\a_{L/R}$. 
However, $Z_L Z_R$ annihilates the R-R
charges (see footnote 3) so one needs to develop a ``picture-raising''
prescription which preserves BRST cohomology. This has been
accomplished in \ref\menew{N. Berkovits, in preparation.}
where it will be shown 
that the R-R charge $C_{(0)}^{\a\b}$
of \char is the $(P_L,P_R)=(-3/2.-1/2)$ version of
the R-R charge $\int d\s \l_L^\a \l_R^\b$
which was proposed in reference \ref\extra{N. Berkovits, `Extra
Dimensions in Superstring Theory', preprint IFT-P.033/97, hep-th
9704109.}
using a twistor-like construction.

\newsec{Supersymmetry Algebra of the Superstring}

Finally, it will be shown that the supersymmetry transformations
of the superstring
close to an N=2 SUSY algebra including R-R central charges.
Since the SUSY generators carry 
picture, one needs to be careful to choose the right picture when 
defining the supersymmetry transformations of the string field.
For a global supersymmetry transformation parameterized by
$u_\a$ and $\tilde u_\a$, the correct choice is
\eqn\susy{\d |\Phi\>_{NS-NS}= \tilde u^\a \bar q^\a_L 
|\Phi\>_{R-NS}
+u^\a q^\a_R
|\Phi\>_{NS-R},}
$$\d |\Phi\>_{R-NS}= \tilde u^\a q^\a_L 
|\Phi\>_{NS-NS}
+u^\a  q^\a_R
|\Phi\>_{R-R},$$
$$\d |\Phi\>_{NS-R}=  \tilde u^\a \bar q^\a_L 
|\Phi\>_{R-R}
+u^\a \bar q^\a_R
|\Phi\>_{NS-NS},$$
$$\d |\Phi\>_{R-R}=  \tilde u^\a q^\a_L 
|\Phi\>_{NS-R}
+u^\a \bar q^\a_R
|\Phi\>_{R-NS},$$
where $q^\a_{L/R}$ carries $-1/2$ picture and $\bar q^\a_{L/R}$ carries
$+1/2$ picture. The easiest way to define the action of $q^\a_{L/R}$ or
$\bar q^\a_{L/R}$ on the string field is to first write the
string field in terms of the SL(2)-invariant vacuum, 
then take the contour integral of $\int d\s 
e^{-\half\phi_{L/R}}\Sigma_{L/R}^\a$
or $\int d\s 
(e^{\half\phi_{L/R}}
\G_\mu^{\a\b}\Sigma_{L/R}^\b
\p_{L/R}x^\mu 
+b\eta e^{{3\over 2}\phi_{L/R}}\Sigma_{L/R}^\a)$ around the vertex operator,
and finally
re-express the resulting vertex operator in terms of the 
string vacuum. Note that in the gauge where $\tilde \phi=M_\mu^\a
=N^\a=A^{\a\b}_{(2)}=0$,
one also needs to include compensating gauge transformations which cancel 
the supersymmetry transformation of 
$\tilde \phi$, $M_\mu^\a$, $N^\a$ and $A_{(2)}^{\a\b}$.
For example, $\d |\Phi\>_{NS-NS}$ needs to include the term 
$Q \tilde u^\a \tilde \tau^\a(x)$
$ c_+^0 \b_L^\half \b_R^\half ~|0\>_{NS-NS}$ to cancel the variation
of $\tilde \phi$, and $\d |\Phi\>_{R-R}$ needs to include the term
$Q u^\b \tilde\Xi^\a(x)$
$c_+^0 (\b_L^0)^2 ~|0\>_{R-R}^{\a\b}$
to cancel the variation of $A_{(2)}^{\a\b}$.

The simplest non-trivial computation is the 
field-independent part of the transformation
resulting from the commutator of a global
supersymmetry transformation with a local supersymmetry transformation.
(The field-independent part vanishes for the
commutator of two global SUSY transformations, and is complicated
for the commutator of two local SUSY transformations.)

Using \susy and \ar, the field-independent part of this commutator is 
\eqn\com{
\d |\Phi\>_{NS-NS}= 
\tilde u^\a \bar q_L^\a ~
Q \tilde \e^\b c_+^0 (\b_L^0 \b_R^\half -
(\b_L^0)^3 \g_R^\half )~
|0\>_{R-NS}^{\b}} 
$$+u^\a q_R^\a ~
Q \e^\b \b_L^\half~ |0\>_{NS-R}^{\b} ~+~
Q \tilde u^\a \p^{\a\g}\tilde \e^\g
c_+^0 \b_L^\half \b_R^\half ~|0\>_{NS-NS} ,$$ 
$$\d |\Phi\>_{NS-R}= 
\d |\Phi\>_{R-NS}=0,$$ 
$$\d |\Phi\>_{R-R}= \tilde u^\a q^\a_L ~
Q \e^\b \b_L^\half ~|0\>_{NS-R}^{\b}
~+u^\a \bar q^\a_R~
Q \tilde \e^\b c_+^0 (\b_L^0 \b_R^\half -
(\b_L^0)^3 \g_R^\half )~
|0\>_{R-NS}^{\b}
$$
$$+~
Q u^\b \p^{\a\g}\tilde\e^\g
c_+^0 (\b_L^0)^2 ~|0\>_{R-R}^{\a\b}, 
$$
where $[u^\a,\tilde u^\a]$ are global parameters and
$[\e^\a(x),\tilde\e^\a(x)]$ are local parameters.

Converting to
the SL(2)-invariant vacuum to compute the action of $q_{L/R}^\a$ and
$\bar q_{L/R}^\a$, and then re-expressing in terms of the original
vacuum, one finds
\eqn\result{
\d |\Phi\>_{NS-NS}= 
\G^\mu_{\a\b}
Q (u^\a \e^\b \b_L^\half \psi_R^{\half \mu}
+\tilde u^\a  \tilde \e^\b \psi_L^{\half\mu} \b_R^\half
+\tilde u^\a \p^{\b\g} \tilde \e^\g 
c_+^0 \b_L^\half \b_R^\half)~
|0\>_{NS-NS},}
$$\d |\Phi\>_{NS-R}= 
\d |\Phi\>_{R-NS}=0,$$ 
$$\d |\Phi\>_{R-R}= Q(\tilde u^\a \e^\b
+ u^\b  \tilde\e^\a) \b_L^0~ 
|0\>_{R-R}^{\a\b}.$$

So the field-independent part of the transformation resulting from 
the commutator of a local and global supersymmetry transformation is
given by
\eqn \last{[ \d_{q_L^\a} (\tilde\e^\a(x))
+ \d_{q_R^\a} (\e^\a(x))~,~
\d_{q_L^\b} (\tilde u^\b)
+ \d_{q_R^\b} (u^\b)]}
$$=
\d_{P_R^\mu} (\e^\a (x) \Gamma^\mu_{\a\b} u^\b)
+ \d_{P_L^\mu} (\tilde \e^\a(x) \Gamma^\mu_{\a\b} \tilde u^\b)
+ \d_{C_{(0)}^{\a\b}} (\tilde u^\a \e^\b(x) +
u^\b \tilde \e^\a(x))$$
where 
$\d_{P_R^\mu}(y^\mu)$ 
and $\d_{P_L^\mu}(\tilde y^\mu)$ 
are NS-NS gauge transformations parameterized by
$y^\mu$ and $\tilde y^\mu$ of equation \ar,
$\d_{q_R^\a} (\e^\a)$ and
$\d_{q_L^\a} (\tilde\e^\a)$
are R-NS and NS-R gauge transformations parameterized by
$\e^\a$ and $\tilde \e^\a$ of equation \ar,
and
$\d_{C_{(0)}^{\a\b}} (\rho^{\a\b}_{(0)})$
are R-R gauge transformations parameterized by
$\rho^{\a\b}_{(0)}$ of equation \th.
Therefore, supersymmetry transformations of the Type II superstring
form an N=2 SUSY algebra with an NS-NS one-brane central charge and
with 256 R-R central charges.

Note that, unlike the conjecture of \extra, 
there is no $e^{\phi}$ dependence in $\{q_L^\a, q_R^\b\}$.
Nevertheless, the claim is still correct in \extra
that momentum in the eleventh direction,
$P_{11}$, should be associated with $e^{\phi}$ times the
R-R central charge $C_{(0)}^{\a\a}$. The $e^{\phi}$ dependence comes
from the fact that the N=1 D=11 SUSY algebra is
\eqn\elev{\{\hat q^A, \hat q^B\}= \G_M^{AB} E^{M m} P_m}
where $A=1$ to 32 are SO(10,1) spinor indices,
$M=0, ..., 9, 11$ are flat vector indices,
$m=0, ..., 9, 11$ are curved vector indices, $\hat q^A$ are
the N=1 D=11 SUSY generators, and $E^{Mm}$ is the
D=11 vierbein.
When the D=10 metric is flat, $E^{Mm}= e^{{1\over 3}
\phi}\d^{Mm}$ for $M$=0 to 9
and $E^{11 ~m}= e^{-{2\over 3}\phi}\d^{11~ m}$.\ref\wit
{E. Witten, Nucl. Phys. B443 (1995) 85.} 
So if $P_m$ is the ten-dimensional momentum for $m=0$ to 9, then
$\hat q^A$ = $e^{{1\over 6}\phi} q_{L}^\a$ for $A=1$ to 16 and
$\hat q^A$ = $e^{{1\over 6}\phi} q_{R}^\a$ for $A=17$ to 32.  
Comparing with \last, this implies that
$P_{11}= {1\over {32}} e^{\phi} C^{\a\a}_{(0)}$.

{\bf Acknowledgements:} I would like to thank I. Bars,
S. Ramgoolam and W. Siegel 
for useful conversations. This
work was financially supported by 
FAPESP grant number 96/05524-0.

\listrefs
\end